# Estimating the Demand Factors and Willingness to Pay for Agricultural Insurance

Osman Gulseven*

Middle East Technical University, Department of Economics, A05, 06800, Ankara, Turkey

*Email of the corresponding author: gulseven@metu.edu.tr

**Abstract**

This article investigates the effect of prices and socio-demographic variables on the farmers' decision to purchase agricultural insurance. A survey has been conducted to 200 farmers most of whom are engaged in diversified income-generating activities. The logistic estimation results suggest that education and household income from farming activities positively affect the likelihood of purchasing insurance. The demand for insurance is negatively correlated with the premium paid per insured value, suggesting that insurance is a normal good. Farmers are willing to pay (WTP) increasingly higher premiums for contracts with higher coverage ratio. According to the valuation model, the WTP declines sharply for coverage ratios under 70%.

## 1. Introduction

It is well known that agriculture is a risky business. Several risks exist from the beginning of the production cycle until the harvest period. While mitigation is the primary choice to reduce undesirable consequences, the realization of losses can be devastating on the farmers' income. Studies suggest that an effective insurance scheme can transfer risk to other financial markets (Joint Research Centre 2013; Bielza, *et al.* 2009; Carriquiri & Osgood 2012). By reducing the production risk, insurance can motivate farmers to specialize in high-productivity farming activities (Barrett & McPeak 2006). It is also possible that the right insurance schemes can motivate farmers to apply effective mitigation techniques (World Bank 2005; Mahul & Stutley 2010).

The trend around the world is to promote and subsidize agricultural insurance as the common practice for optimal risk management. In theory, the difference between farmers willingness to pay (WTP) for insurance and the actuarially fair premium needs to be calculated and then financed by the state (Juanchang & Jiyu 2010; Herbold 2010). In practice, the decisions are usually not based on economic analysis, but rather political motivations. The recent US Farm Bill 2014 shifted the direct payments into crop insurance subsidies. This bold move created additional controversy around why insurance needs to be subsidized and who benefits from the subsidized insurance policies. According to Lynch & Bjerga (2013), a significant portion of those subsidies go to insurance companies. Their report also suggests that affluent growers benefit more under the new subsidized insurance schemes.

In Europe, the agricultural insurance system is widely diversified. While European Commission aims to unify the different schemes, each country has its own system. Some member states such as Spain offer substantial subsidies, whereas some states do not have such advanced subsidization schemes (Anton & Kimura 2011). The rules and regulations also differ within each state. Public insurance is the dominant factor in Spain, whereas farmers enter into collective bargaining in Italy.

The Turkish agricultural insurance system is somewhere in between the European and American systems. While self-sustainability is the primary concern, it is the government policy to promote and subsidize agricultural insurance. The current norm in Turkey is that about 50% of the premium is paid by the government, whereas the rest is paid by the farmer. For some commercially vital products that have export potential, the subsidies can be as high as 66%. The problem with those subsidies is that they can cause an imbalance not only in the insurance market but also the agrarian product market as they are not based on any economic model.

A better way to determine the optimal subsidy rate is to derive the actual WTP for insurance and the state shall pay the rest. In order to understand how much farmers are willing to pay for agricultural insurance, a survey has been conducted in the central Anatolian basin in Turkey. While the survey area is a localized, the results and the methodology can easily be extended to different regions of the world.

## 2. Data

The data is based on a recent survey conducted in capital region of Turkey. 200 farmers from the region were interviewed about their opinions on potential insurance schemes. Out of 200 observations 128 has purchased insurance, whereas 72 is not insured. Most of these farmers are also engaged in other income-generating activities. The annual household income from farming activities is relatively low, but this is offset by other income. These farmers adopt self-sufficiency concept. Also, most of their farming activities are off the book. A substantial portion of the farm-produced products are utilized for family use without making into the commercial markets.

Insurance is a very familiar concept among these farmers. Thanks to extensive government-supported ad campaigns, the link between insurance and access to credit is well known in the area. Many have to purchase insurance to get access to credit. In fact, it is the primary reason for insurance coverage. Of course there are reasons as well. The responses of farmers, regarding their decision to buy agricultural insurance are listed in Table 1.

**Table 1. Why Farmers Buy Insurance?**

| Reason | Percentage Frequency | Actual Frequency |
| --- | --- | --- |
| Legal Requirement to Access Credit | 40.625% | 52 |
| Protection Against Losses | 21.875% | 28 |
| Networking Affects | 18.75% | 24 |
| Government Ads | 12.5% | 16 |
| Other / No Response | 6.25% | 8 |

As a part of the survey, we asked about the primary reasons to purchase insurance. The responses revealed interesting results. The number one reason why farmers prefer to buy insurance is because they have to. 40.625% of the respondents indicated that it is legally required to be covered under an insurance scheme. This is quite interesting as the agricultural insurance is not mandatory in the area. However, it is linked with credit. So, insurance provides access to credit. The bundling of insurance with credit is the primary reason for insurance purchase.

Protection is the primary reason for insurance purchase among only 21.875% of the respondents. This low ratio is probably due to the fact that farmers are engaged in diversified income activities. The diversification of income sources offers the farmers a relatively stable income, reducing the need for additional income protection measures.

18.75% of the farmers indicated that they are purchasing insurance because their neighbors/relatives are also buying it. Government sponsored ads are the primary purchase reasons for 12.5% of the farmers.

**Table 2. Why Farmers Do Not Buy Insurance?**

| Reason | Percentage Frequency | Actual Frequency |
| --- | --- | --- |
| No need | 44.44% | 32 |
| Cost issues | 33.33% | 24 |
| Too small to be insured | 16.67% | 12 |
| Other / No Response | 5.56% | 4 |

About 36% of those surveyed responded that they are not currently insured. A dominant portion (44.44%) of the respondents who are not insured stated that they believe agricultural insurance is unnecessary. Those farmers have an idea about agricultural insurance, but they do not like the fact that their premiums are not returned if the risk is not realized. 33.33% of those who prefer to be uninsured listed the additional cost as their primary reason not to buy insurance. About 16.67% of the farmers suggested that they are small farmers and that is the reason why they are not insured.

**3. Methods**

The methodology used in this article is two-fold. First, a logistic regression is utilized to quantify the factors which might affect the agricultural households' demand for insurance. Following Train's (2003) notation and McFadden's framework (1980) the estimated regression is parameterized as

$$P(U_i > U_j, \forall i \neq j) = \frac{\exp(V_i)}{\sum_j \exp(V_j)} \quad P(U_i > U_j, \forall i \neq j) = \frac{\exp(V_i)}{\sum_j \exp(V_j)} \quad (1)$$

If we define the related parameters in linear form such that $V_{nj} = \beta x_{nj}$, then the logistic probability with only two alternatives can be written as

$$P_n = \frac{e^{\beta' x_n}}{1+e^{\beta' x_n}} \quad P_n = \frac{e^{\beta' x_n}}{1+e^{\beta' x_n}} \tag{2}$$

In the above regression $x_n$ refers to household specific socio-demographic information such as education level, income from farming activities, income from off-farming activities, household size, and union membership. The vector $\beta$ measures the effects of the above-mentioned variables on insurance purchase probabilities.

Factor Effect = $\beta_1$ (Education) + $\beta_2$ (Farming Income) + $\beta_3$ (Off-Farm Income) + $\beta_4$ (Household Size) + $\beta_5$ (Union Membership)  (3)

Next, I estimated the farmer's WTP for insurance using contingent valuation method. Contingent valuation models and willingness to pay studies have been widely utilized in designing optimal mechanism designs (Horowitz & McConnell 2002). Several methods were suggested on this technique in Alberini & Kahn (2006). Following their work, both open-ended and take-it or leave-it types of questions were used to derive the demand curves. In the derived demand curves, the prices are defined as the percentages of insured values and the quantities are defined as percentages of farmers who are willing to pay those prices. However, some farmers responded to low-coverage insurance products with a valuation of "0". The zero values were dealt as suggested by Strazzera *et al.* (2003).

**4. Binary Logistic Results**

In the binary logistic regression, success is defined as insurance purchase, whereas failure is defined as not being covered. Thus, the dependent variable takes '0' for the null case. The explanatory variables are education, farming income, off-the farm income, household size and experience with union. The results are as follows.

**Table 3. Logistic Regression Results**

| Predictor | Coefficient | Standard Deviation | *p-value* | Odds Ratio | Lower Confidence | Upper Confidence |
|---|---|---|---|---|---|---|
| Constant | -1.177 | 0.637 | 0.065 | | | |
| Education | 0.612* | 0.252 | 0.015 | 1.84 | 1.12 | 3.03 |
| Farming Income | 0.126* | 0.057 | 0.026 | 1.22 | 1.06 | 2.42 |
| Off-Farm Income | 0.036 | 0.056 | 0.516 | 1.02 | 0.81 | 1.31 |
| Household Size | -0.115 | 0.209 | 0.582 | 0.89 | 0.59 | 1.34 |
| Union Membership | 0.389 | 0.382 | 0.308 | 1.48 | 0.7 | 3.12 |

* Significant at 5%. Log-Likelihood = -104.582

The *p-value* for the regression suggests a significant model. Chi-square tests measured as suggested by Pearson, Deviance, and Hosmer-Lemeshow are 175, 187, and 21. Thus, the defined equation works well with the data. The observed frequencies and expected frequencies also match closely with each other. The concordant pair association is calculated as 75.5%, and the discordant pair association is 23.9%. These ratios suggest strong measures of association between the response variables and predicted probabilities.

Education, which is defined as a scale from 0 to 4, is a strong factor in explaining the decision to purchase insurance. Its *p-value* is 0.015, which suggests a statistically significant variable. Both farming income and off-the-farm income are also positively related with the insurance purchase decision, but only farming income is statistically significant. Household size and union membership are not statistically significant factors.

**5. Estimating Willingness to Pay for Insurance**

Using contingent valuation model, I estimated the farmers' demand function for a range of premium prices which are defined in terms of insured values. In the survey, there were specifically designed questions on estimating how much farmers are willing to pay for different levels of insurance coverage. This implicit insurance pricing model is derived for crop insurance, fruit insurance, and livestock insurance.

*4.1 Estimating Willingness to Pay for Crop Insurance*

Crops are usually the easiest agricultural products to be insured. As they do not suffer that much from unexpected rain, frost, or similar risks, they are subject to relatively lower risks compared to fruits and livestock.

The following graph visualizes the demand curves for crop insurance at each level of coverage:

**Graph 1: Demand for Crop Insurance**

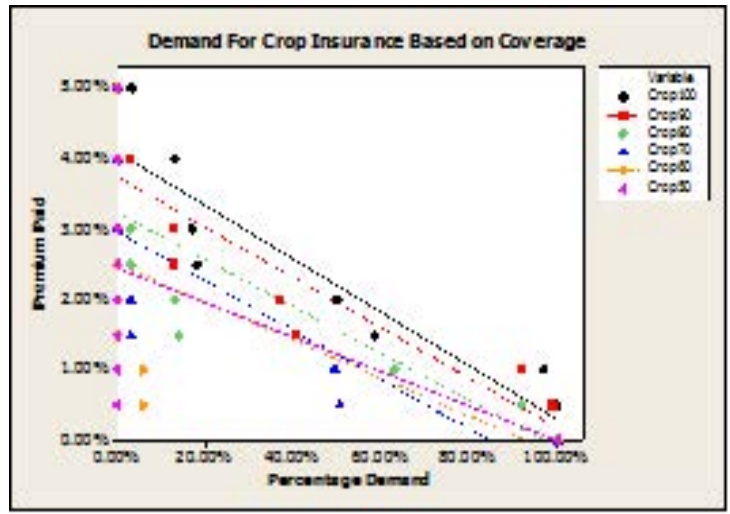

The regression line on the top of the graph gives the demand curve for full coverage with no deductibles. This graph suggests that the maximum willingness to pay for crop-value insurance is 5% of the premium for the insurance type which offer full coverage with no deductibles. However, only 3% of the farmers are willing to pay a premium up to that much. In order to cover at least half of the population, the maximum premium should be 2% of the insured value for crop insurance for such coverage. If the premiums fall to 1%, then almost 97% of the farming population will show interest in full crop insurance. As the level of coverage declines, farmer's willingness to pay also declines. The decline in willingness to pay becomes much sharper when the coverage ratio falls below 70%.

*4.2 Estimating Willingness to Pay for Fruit Insurance*

The willingness to pay for fruit insurance is higher than that of crop insurance. The graph below shows the farmers' hypothetical demand curve for fruit insurance. Similarly, the price premium in y-axis is defined in terms of insured value, and the x-scale shows percentage of farmers showing purchase interest at each premium.

**Graph 2: Demand for Fruit Insurance**

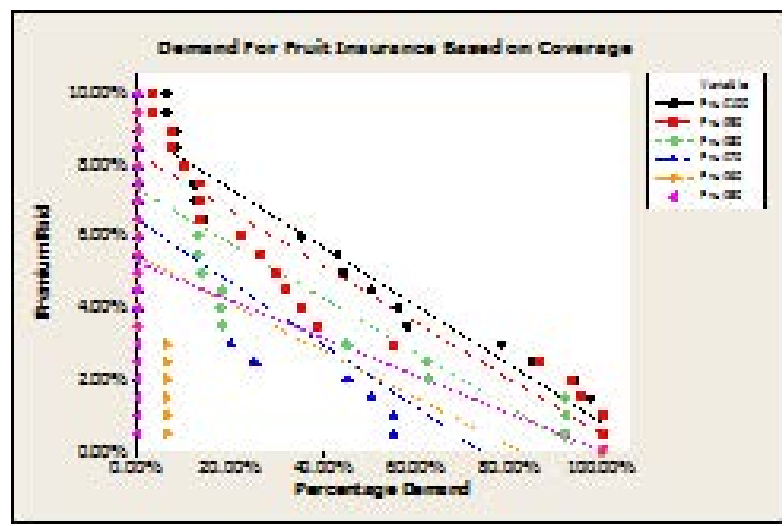

Fruit is a more valuable and vulnerable asset compared to crop. The revenue per fruit land is substantially higher than the revenue derived from crops for the same area. Fruit production is also riskier. While frost and hail does not have that much affect on crop production, an unexpected realization of these risks can be devastating for fruit producers. Aware of those risks, the farmers' willingness to pay for fruit insurance is substantially higher. Some farmers are willing to pay for as high as 10% of the insured value for that type of insurance. A premium of 4% covers 56% of the farmers who are protected by full coverage. The participation ratio falls to 35% for 90% coverage (10% deductible), and sharply falls to 18% for 80% coverage (20% deductible).

*4.3 Estimating Willingness to Pay for Livestock Insurance*

Insuring livestock has similarities and also differences from insuring vegetations. Livestock is usually well protected from rain and frost. Farmers can also take precautionary actions against drought as well. The main risk factor in livestock is the spread of disease. The appearance of diseases can be both sporadic and systematic. Once a contagious disease emerges in a barn, there is a high risk of losing the entire pack. Therefore, it is of critical importance for the farmers to have insurance to protect themselves from such disasters.

By quantifying the respondents' answers, one can drive the demand function for insurance. Similar to crop and fruit insurance, I derived the insurance demand for full coverage, 90% coverage, 80% coverage, 70% coverage, 60% coverage and 50% coverage levels. The graph below visualizes the demand for livestock insurance at each level of coverage.

**Graph 3: Demand for Livestock Insurance**

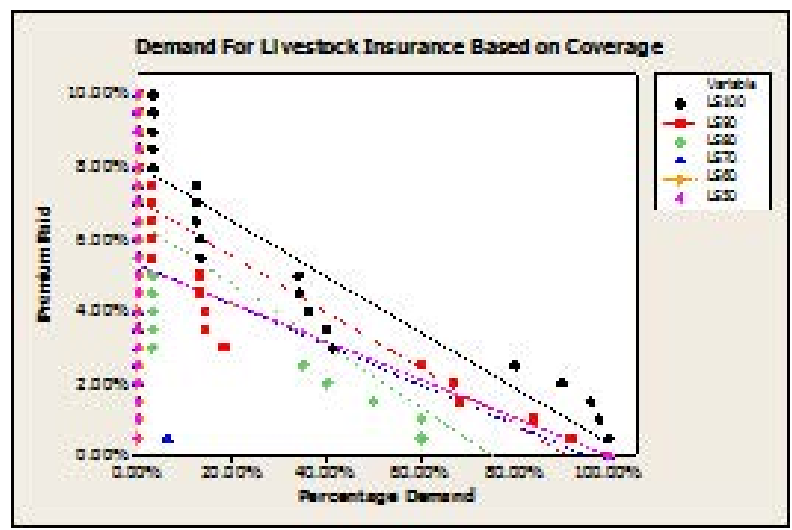

Similar to the case for fruits, farmers are willing to pay pretty high premiums for livestock insurance compared to crop insurance. 3% of the farmers indicated that they are willing to pay up to 10% for full coverage. A modest premium of 2.5% for full coverage is enough to provide insurance to 80% of the respondents. This ratio falls to 60% for a coverage ratio of 90%, and it falls to 35% for a coverage ratio of 80%. Farmers do not show any interest in purchasing insurance for livestock that offers coverage of below 70% of insured value.

**5. Conclusion**

The analysis of survey responses revealed several interesting results. According to the survey, farmers are buying insurance primarily because it is attached to credit. Government sponsored ads and networking effects are also positive impediments to acquire insurance coverage. Education and farming income increase the chances to buy agricultural insurance among rural households. Similar to Sundar and Ramakrishnan's findings (2013), a substantial portion of the farmers believe that insurance is not necessary for small farmers. This belief is a serious drawback in insuring small farmers.

Horowitz & McConnell suggest that the difference between willingness to pay and willingness to accept also depends on income factor (2003). As income rises, this price differential might get narrower reducing the need for insurance subsidies. The demand for insurance is a normal good and the farmers

WTP declines sharply for lower coverage levels. The results suggest that most farmers are willing to pay a meaningful amount for full coverage, but their WTP declines sharply for insurance schemes that offer less than 80% coverage. It is also striking that most farmers are not willing to pay anything for coverage below 70%. That creates a substantial challenge to index-based insurance schemes where the payments are determined by an external index threshold. It is suggested that the correlation between the index and the losses should be at least 70% for a sustainable index-insurance scheme.